\documentclass[12pt]{article}

\usepackage{scicite}
\usepackage[dvipdfmx]{graphicx}
\usepackage{float}
\usepackage{times}
\usepackage{color}

\newenvironment{sciabstract}{%
\begin{quote} \bf}
{\end{quote}}

\title{Dominant non-local superconducting proximity effect due to electron-electron interaction in a ballistic double nanowire}

\author
{Kento Ueda,$^{1\ast \dagger }$ Sadashige Matsuo,$^{1 \ast \dagger }$\\ Hiroshi Kamata,$^{1,2}$ Shoji Baba,$^{1}$ Yosuke Sato,$^{1}$ Yuusuke Takeshige,$^{1}$\\ Kan Li,$^{3}$ S${\rm \ddot{o}}$ren Jeppesen,$^{4}$ Lars Samuelson,$^{4}$ Hongqi Xu,$^{3,4\ast}$\\ and Seigo Tarucha$^{1,2\ast}$\\
\\
\normalsize{$^{1}$Department of Applied Physics, University of Tokyo,}\\
\normalsize{7-3-1 Hongo, Bunkyo-ku, Tokyo 113-8656, Japan}\\
\normalsize{$^{2}$Center for Emergent Matter Science, RIKEN,}\\ 
\normalsize{2-1 Hirosawa, Wako-shi, Saitama 351-0198, Japan}\\
\normalsize{$^{3}$Beijing Key Laboratory of Quantum Devices,}\\
\normalsize{Key Laboratory for the Physics and Chemistry of Nanodevices}\\
\normalsize{and Department of Electronics, Peking University, Beijing 100871, China}\\
\normalsize{$^{4}$Division of Solid State Physics/NanoLund, Lund University, box 118, SE-221 00 Lund, Sweden}\\
\\
\normalsize{$^\ast$To whom correspondence should be addressed;}\\
\normalsize{E-mail: ueda@meso.t.u-tokyo.ac.jp, matsuo@ap.t.u-tokyo.ac.jp,}\\
\normalsize{hqxu@pku.edu.cn, tarucha@ap.t.u-tokyo.ac.jp.}\\
\normalsize{$^\dagger $These authors contributed equally to this work}
}

\date{}

\begin{document} 


\baselineskip24pt


\maketitle


\begin{sciabstract}
Cooper pair splitting (CPS) can induce non-local correlation between two normal conductors that are coupled to a superconductor.
CPS into a double one-dimensional electron gas is an appropriate platform for extracting a large number of entangled electron pairs and one of the key ingredients for engineering Majorana fermions with no magnetic field. 
In this study, we investigated CPS using a Josephson junction of a gate-tunable ballistic InAs double nanowire. 
The measured switching current into the two nanowires is significantly larger than the sum of the switching current into the respective nanowires, indicating that the inter-wire superconductivity is dominant compared to the intra-wire superconductivity. 
From its dependence on the number of propagating channels in the nanowires, the observed CPS is assigned to one-dimensional electron-electron interaction. 
Our results will pave the way for the utilization of one-dimensional electron-electron interaction to reveal the physics of high-efficiency CPS and to engineer Majorana fermions in double nanowire systems via CPS.
\end{sciabstract}


\section*{Introduction}
The superconducting proximity effect can induce superconducting correlation in normal conductor nanostructures in contact with a superconductor~\cite{proximity}, and it can therefore provide a platform for the engineering of exotic phenomena and novel superconductivity. 
When two normal conductors are closely spaced, the proximity effect can inject non-local electron correlation between the two normal conductors, which is referred to as Cooper pair splitting (CPS)~\cite{recherprb2001,lesovikepj2001,bouchiatnanotech2002,benaprl2002,recherprb2002,chtchelkatchevprb2002,recherprl2003,bursetprb2011}.
CPS has been intensively studied in terms of generating non-locally entangled spin pairs for applications in quantum information techniques. As a result, to date, CPS experiments have been exclusively performed on superconductors in contact with two quantum dots~\cite{hofstetternature2009,herrmannprl2010,schindeleprl2012,dasnatcommun2012,tanprl2015, russellnatcommun2015,fulopprl2015,ivanscirep2016,babanjp2018}.
In such devices, local pair tunneling (LPT), which occurs when two electrons in a single Cooper pair tunnel into the same dot, is strongly suppressed due to the large cost of electrostatic energy, and CPS can be dominant over LPT. 
To acquire a high-efficiency CPS to LPT ratio in the system, it is necessary to reduce the amount of dot to superconductor tunnel coupling and/or increase the electrostatic energy of the dots~\cite{recherprb2001}.
However, in reality, this will also lead to a significant reduction of the conductance of the dots. 
Then, signals of CPS become tiny as the CPS to LPT ratio increases. 
Therefore, this way of engineering CPS is inappropriate for further experimental studies on the nature of split electrons~\cite{brauneckerprl2013}.

One-dimensional (1D) electron gases can be considered as an alternative to quantum dots in the study of CPS because the 1D repulsive electron-electron (e-e) interaction is sporadically screened out~\cite{tomonaga,luttinger}, which can suppress LPT. 
Theoretically, the physics has already been developed for a hybrid system of two Tomonaga-Luttinger liquids (TLL) contacted to a superconductor~\cite{recherprb2002}. Unlike the case for quantum dots, the CPS efficiency is only affected by the 1D e-e interaction and not by the tunnel coupling of the 1D electron gas to the superconductors. This difference is crucial for extracting a large number of entangled spin pairs and holding non-local superconducting correlation into the double 1D electron gas. The former is useful for implementing efficient entangled spin pair sources in solid-state systems. 
For the latter, when the inter-wire proximity-induced superconductivity via CPS is dominant over the intra-wire superconductivity via LPT in a parallel double nanowire (DNW), such as InAs or InSb, with strong spin-orbit interaction in terms of superconducting gap energy, the system is predicted to indicate time-reversal invariant topological superconductivity in which Kramers pairs of Majorana fermions (MFs) appear at the edges~\cite{jelenaprl2014,jelenaprb2014,ebisuptep2016,schradeprb2017,reegprb2017,thakurathiprb2018}. 
Some signatures of topological superconductivity and MFs~\cite{kitaev,hasanrmp2010,qirmp2011} have recently been reported in a single NW contacted to a superconductor~\cite{mourikscience2012,rokhinsonnatphys2012,dengnl2012,dasnatphys2012,dengscirep2014,albrechtnat2016,dengscience2016, tiiranatcommun2017}. 
However, in such a device, a strong magnetic field is required to realize the MFs, which can affect the robustness of MFs due to the quasi-particle~\cite{rainisprb2012}. On the other hand, the proximity-induced superconductivity in the double nanowire can remove the restriction of a strong magnetic field and become a more robust platform for MFs and future topological quantum circuits. 
However, high-efficiency CPS and dominant inter-wire superconductivity have never been demonstrated via the CPS in 1D electron gases.

In this work, we report the first observation of CPS in a ballistic DNW Josephson junction based on the measurement of switching current. The CPS contribution in the switching current is dominant over the LPT contribution, which is ascribed to the 1D e-e interaction effects. 
In addition, Josephson junction devices enable us to evaluate the superconducting gap energies of the inter-wire and intra-wire superconductivity as products of the switching current and normal resistance. We found that the gap of the inter-wire superconductivity is larger than that of the intra-wire superconductivity, which is one of the necessary conditions to realize time-reversal invariant topological superconductivity and Kramers pairs of MFs. 
Our results will pave the way for the utilization of the 1D e-e interaction to reveal the physics of high-efficiency CPS and to engineer topological superconductivity and MFs in a double nanowire of not only InAs or InSb~\cite{jelenaprl2014, jelenaprb2014} but also chiral and helical edge states of topological insulators~\cite{clarkenatcommun2013, mongprx2014, clarkenatphys2014}.

\section*{Josephson junction device of a double InAs nanowire}
The Josephson junction device used in this study has an InAs DNW between two Al electrodes with a small separation of 20 nm. 
The InAs NWs have a diameter of approximately 80 nm and are grown by chemical beam epitaxy. We transferred the NWs on the growth substrate on to an Si substrate covered by a 280 nm thick ${\rm SiO_2}$ film and picked out closely spaced parallel DNWs to make Al-DNW-Al junctions. We made a polymethyl methacrylate  pattern of the Al electrodes using electron beam lithography and performed a NW surface treatment before evaporating Ti (1 nm)/Al (100 nm): reactive ion etching to remove the polymethyl methacrylate residue and sulfur passivation to remove the surface oxide (see Supplementary Note 1 and Supplementary Figs. S1 and S2). 
A scanning electron microscope image and schematic of the device are shown in Figs. 1(a) and (b), respectively. The electron conductions of the two NWs are independently modulated using two separate gate electrodes labeled g1 and g2 (orange). 
In this paper, we refer to the nanowire closest to g1 (g2) as NW1 (NW2).
For the fabrication of the gate structure, we grew a 40 nm thick ${\rm Al_2O_3}$ layer by atomic layer deposition and fabricated the gate electrodes of Ti (5~nm)/Au (150~nm).

\section*{Normal conductance in a double InAs nanowire}
First, we measured the differential conductance of the Josephson junction device at 50 mK under a larger magnetic field than the critical field for the Al electrode to characterize the normal transport property of the DNW (see Supplementary Note 2 and Supplementary Fig. S3). 
For the electron transport measurement, we used a standard lock-in technique.
Figure 1(c) shows the measured differential conductance $G$ as a function of the two voltages $V_{g1}$ and $V_{g2}$ for g1 and g2, respectively. 
The pinch-off regions for NW1 and NW2 are located below the blue solid line and to the left of the red solid line, respectively. 
The conduction can be divided into four regions separated by the red and blue solid lines: conduction of NW1 only (upper left), NW2 only (lower right), both NWs (DNW) (upper right), and no conduction (lower left). 

Figure 1(d) shows the conductance line profiles related to Fig. 1(c). 
The blue (red) lines indicate $G$ of the only NW1 (NW2) measured by setting $V_{g2}$ between -5.0 and -8.0 V ($V_{g1}$ between -17.0 and -20.0 V). 
For both conductance lines, we observe plateau-like structures characterized by the quantized conductance of $G=me^2/h$ with $m$ (= 2, 4, 6). The typical conductance data are shown by the bold lines. Fluctuations of the conductance are likely due to impurity scattering, which depends on the two gate voltages. From the observation of the conductance plateau-like features, we confirm ballistic transport in each NW. Moreover, no definite tunnel junctions formed at the interface of the Al-NW junctions. 
In Fig. 1(c), the dashed lines parallel to the blue or red solid lines or connecting the onsets of the respective conductance plateaus mark the transitions between neighboring plateaus in each NW. 
The conductance in each region bounded by two sets of neighboring dashed lines is then given by $G(m,n) = me^2/h+ne^2/h$, where $m$ and $n$ denote the number of propagating 1D channels in NW1 and NW2, respectively, and we call this region ($m,n$).

Finally, to characterize the electron transport in the DNW region, we show the conductance line profile along the thick purple line in Fig. 1(c), as plotted in Fig. 1(e). This purple line crosses (0,0) to (2,2) and to (4,4). Correspondingly, we observed conductance plateaus of 4 and 8 $e^2/h$. Hence, we confirmed that DNWs are ballistic, and the conductance in the normal state can be understood as the sum of the conductance values in the separate NWs.
\section*{Supercurrent in the double nanowire Josephson junction}
Next, we measured the differential resistance $R$ against bias current $I$ under the magnetic field $B = 0$ T to observe the supercurrent. 
Figures 2(a) and (c) present typical results measured in (2,0), (4,0), and (6,0) of the NW1 region and in (0,2), (0,4), and (0,6) of the NW2 region, respectively. For sweeping the current from positive to negative, $R$ becomes almost zero in the finite current range centered at I = 0 A, which indicates a supercurrent flowing through NW1 or NW2, and then abruptly increases and comes to a peak. We determine $I_{sw}$ at the peak position. Almost the same $I_{sw}$ is derived from the peak in the positive current region. We measured $I_{sw}$ at several points on the same plateaus and took the average (see Supplementary Note 4 and Supplementary Fig. S5). The $I_{sw}$ against $G$ plot shown in Figs. 2(a) and (c) are plotted in Figs. 2(b) and (d), respectively. Clearly, $I_{sw}$ monotonically increases with $G$. Here, we use the $G$ shown in Fig. 1(c). 
We note that the Josephson junction is ballistic. 
This is supported by the normal conductance results, as discussed above, and the observed multiple Andreev reflection (see Supplementary Note 3 and Supplementary Fig. S4)

Next, we measured $R$ against $I$ in (2,2), (4,2), (6,2), (2,4), (2,6), and (4,4) of the DNW regions to study the CPS contributions. Figure 3(a) shows a typical result obtained in (2,2) (black), together with typical results in (2,0) and (0,2) (blue and red, respectively). Similarly, Fig. 3(b) shows the result in (4,4) (black), together with typical results in (4,0) and (0,4) (blue and red, respectively). The derived $I_{sw}(2,2) = 11.3$ nA is much larger than $I_{sw}(2,0)+I_{sw}(0,2)=4.78$ nA. Here, $I_{sw}(m,n)$ and $G(m,n)$ are $I_{sw}$ and $G$, respectively, measured in the $(m,n)$ regions. Figure 3(c) shows a plot of $I_{sw}(m,n)$ against $G(m,n)$ in the DNW region (purple triangles) and $I_{sw}(m,0)+I_{sw}(0,n)$ against $G(m,0)+G(0,n)$ for the sum of $I_{sw}$ measured in the respective NW regions (pink circles). $I_{sw}(m,n)$ is explicitly larger than $I_{sw}(m,0)+I_{sw}(0,n)$ for all values of $m$ and $n$.

The switching current $I_{sw}$ for the respective NWs is the contribution from the LPT to the supercurrent. 
On the other hand, the CPS contribution, which only appears when both NWs have finite propagating channels, is observed as the surplus $I_{sw}$ for the DNW compared to the sum of $I_{sw}$ measured for the respective NWs. 
This evaluation method of CPS and LPT has been utilized in studies of CPS of double quantum dots coupled to a superconductor~\cite{recherprb2001, russellnatcommun2015, babaapl2015}.
Therefore, from the large $I_{sw}$ enhancement in the DNW regions shown in Fig. 3(c), it is concluded that there are significant CPS contributions. 
This CPS contribution is described as supercurrent coherently carried by the split electrons in the DNW from one Al contact which are recombined into a Cooper pair at another Al contact.
We note that our device has no quantum dots and the observed large CPS does not originate from the electrostatic energy in the dots, as reported previously. 

\section*{CPS efficiency}
From the $I_{sw}$ results, we evaluated the CPS efficiency $\eta$ defined by 
\[
\eta (m,n)=\frac{I_{sw}(m,n)-(I_{sw}(m,0)+I_{sw}(0,n))}{I_{sw}(m,n)} \times 100 \%.
\]
The calculated values of $\eta (m,n)$ are summarized in Fig. 4(a): 57.3 \%, 31.6 \%, 27.8 \%, 48.8 \%, 41.7 \%, and 47.4 \% for the $(2,2), (4,2), (6,2), (2,4), (2,6),$ and $(4,4)$ regions, respectively. In particular, $\eta (2,2)$ exceeds 50 \%, indicating that CPS rather than LPT is dominant in the supercurrent flowing through the DNW. 
Here, we measured the supercurrent in the Josephson junction; therefore, the two split electrons in the measured CPS component should maintain the singlet-pairing phase coherence, and no contribution from quasi-particle tunneling is included in the results because $I_{sw}$ is unaffected by quasi-particle tunneling.

As a result, $\eta >50$ \% is obtained for (2,2). 
This result clearly indicates that the LPT is significantly suppressed because if no e-e interaction is present in either NW, there is no suppression of the LPT, resulting in $\eta \leq 50$ \%.
This property is also found in the theoretical model of CPS in a junction of a superconductor and a double TLL~\cite{recherprb2002}. 
In this model, no e-e interaction represented as the TLL parameter $K_c=1$ gives $\eta \leq 50$ \% because there is no priority between the CPS and LPT. 
On the other hand, in the case of a finite e-e interaction in each NW with $K_c < 1$, the LPT is at a significant disadvantage to the CPS. 
Our result of $\eta (2,2)=57.3$ \% means that e-e interactions are of importance in the mechanism of the observed CPS.
We note that the small effective electron mass in InAs wires can make the 1D electron-electron interaction more significant and give a smaller value of $K_c$~\cite{hevroniprb2016,satoarxiv2018}.

In Fig. 4 (a), we recognize two additional important features. First, $\eta$ is asymmetric with respect to $m$ and $n$, i.e., $\eta (4,2)<\eta (2,4)$ and $\eta (6,2)<\eta (2,6)$, although the normal state conductance $G$ is the same. 
Also, the first feature is assigned to the asymmetry of the carrier density between the two NWs in the proximity region. 
In the device photograph in Fig. 1(a), we see that NW1 is fully covered by the Al electrodes but NW2 is not; therefore, the NW1 proximity region can have larger carrier density than the NW2 proximity region. 
Indeed, the NW1 pinched-off voltage at $V_{g2} = 0$ V is $V_{g1}\simeq -12$ V, whereas that of NW2 at $V_{g1}=0$ V is $V_{g2}\simeq -2$ V. 
In the theoretical model~\cite{recherprb2002}, lower carrier density in the NWs gives smaller $K_c$ in the TLL case, which means stronger e-e interaction compared to the kinetic energy~\cite{kaneprl1992}, so a larger $\eta$ is expected. 
The stronger e-e interaction and more significant LPT suppression expected in NW2 over NW1 could be the reason for the asymmetry of $\eta (4,2)<\eta (2,4)$ and $\eta (6,2)<\eta (2,6)$.

The second feature is that $\eta$ decreases with increased NW channels. 
This feature is also assigned to the weaker e-e interaction in higher carrier density in the respective NWs because smaller $\eta$ is obtained for plateaus associated with more channels.
Here, we assume that the e-e interaction in the NWs varies with $V_{g1}$ and $V_{g2}$ to be biased towards the NWs located between the two Al electrodes. 
However, the NW length defined by the Al electrode gap is 20 nm, which is comparable to the Fermi wavelength and may be too short to significantly affect the interaction strength. 
We assign this contradiction to a broad potential landscape along the NWs across the boundary with the Al metal. Thus, the electrostatic potential of the gated NW gradually changes to that of the proximity region over a distance much longer than the Fermi wavelength. 
Then, the carrier density and therefore the e-e interaction are tuned by $V_{g1}$ and $V_{g2}$.
Two electrons that split from a Cooper pair in that proximity region propagate through the DNW, generating the CPS supercurrent. 
Hence, the large/small relation of the CPS and LPT, namely $\eta$, changes with $V_{g1}$ and $V_{g2}$, as expected when the 1D e-e interaction is gate-tuned. 

\section*{Gap energy of proximity-induced superconductivity via CPS and LPT}
It is of importance to determine which of the contributions to the proximity-induced superconductivity (CPS or LPT) is larger in terms of the superconducting gap energy.
We define $\xi (m,n) = \Delta_{CPS}(m,n)/\sqrt{\Delta_{NW1}(m,0)\Delta_{NW2}(0,n)}$ to discuss the CPS and LPT contributions, namely the inter-wire and intra-wire superconductivity.
$\Delta_{CPS}(m,n)$, $\Delta_{NW1}(m,0)$, ${\rm and}$ $\Delta_{NW2}(0,n)$ are the superconducting gap energies of the inter-wire superconductivity via CPS, the intra-wire superconductivity via LPT in NW1, and that in NW2, respectively.
$\xi$ is the gap energy ratio between the inter-wire and intra-wire superconductivity; it is a measure for characterizing the topological transition in DNW. The condition of $\xi >1$ should be satisfied for realization of Majorana Kramers pairs in DNWs with no magnetic field~\cite{jelenaprb2014, schradeprb2017}.

In the short ballistic Josephson junction with normal resistance $R_n$, $R_nI_{sw} = \pi \Delta /e$ with a superconducting gap energy of $\Delta$ and elementary charge of $e$~\cite{beenakkerprl1991,furusakiprb1992}.
As shown in Figs.1(c),(d), and (e), the present junction is in the ballistic regime, so $\Delta_{CPS}(m,n)$, $\Delta_{NW1}(m,0)$, ${\rm and}$ $\Delta_{NW2}(0,n)$ can be approximately estimated as $G(m,n)^{-1}(I_{sw}(m,n)-I_{sw}(m,0)-I_{sw}(0,n))$, \\
$G(m,0)^{-1}I_{sw}(m,0)$ and $G(0,n)^{-1}I_{sw}(0,n)$, respectively.
The estimated values of $\Delta_{NW1}$, $\Delta_{NW2}$, $\Delta_{CPS}$ and $\xi$  are summarized in Fig.4(b).
As $m$ or $n$ decreases, $\Delta_{NW1}(m,0)$ and $\Delta_{NW2}(0,n)$ decrease. 
This behavior is consistent with our assumption that LPT is more strongly suppressed for the narrower channel due to the stronger e-e interaction, as discussed above.  
On the other hand, $\Delta_{CPS}$ does not change much, and therefore $\xi$ is larger for the narrower channel. 
As a result, we find $\xi $ larger than unity for (2,2) and (2,4).
Therefore, the necessary condition for topological transition~\cite{jelenaprl2014,jelenaprb2014} is satisfied in our DNW junction.

\section*{Magnetic field dependence}
Finally, we studied the magnetic field dependence of the CPS. Figure 5 shows $I_{sw}(2,2)$ and $I_{sw}(2,0) + I_{sw}(0,2)$ measured under various magnetic fields. It is apparent that $I_{sw}(2,2)$ gradually decreases as the field initially increases up to  $B = 80$ mT, whereas $I_{sw}(2,0) + I_{sw}(0,2)$ is almost unchanged. They become almost identical at $B = 80$ mT and then gradually decrease to zero in the same manner as the field increases up to $160$ mT. This indicates that both $I_{sw}(2,2)$ and $I_{sw}(2,0) + I_{sw}(0,2)$ only depend LPT in NW1 and NW2 for $80~{\rm mT}\leq B\leq 160~{\rm mT}$. Therefore, the CPS contribution is only present in the range of $B = 0$ to $80$ mT, as indicated by the purple-shaded area, whereas the LPT contribution remains substantial. 
Note that essentially identical behavior is observed for $I_{sw}(4,4)$ and $I_{sw}(4,0) + I_{sw}(0,4)$, suggesting that the CPS mechanism is universal for one-dimensional electron systems (see Supplementary Note 5 and Supplementary Fig. S6). 

This peculiar magnetic field dependence is qualitatively explained by means of the critical field $B_c$ of superconducting thin film whose coherence length and penetration length are sufficiently longer than the film thickness $d$~\cite{GL, harperPR1968}.
The theoretical indication of $B_c\approx 1/d$ means that the critical field becomes half as the thin film thickness becomes twice. 
This can be applied for the superconducting wire when the magnetic field is perpendicular  to the wire direction. 
As a result, when the Cooper pair spreads over both NWs in the CPS contribution, the critical field becomes half comparing to that for the LPT contribution in which the Cooper pairs are localized in either of the NWs. 
Therefore, the magnetic field dependence observed for $I_{sw}(2,2)$ and $I_{sw}(2,0) + I_{sw}(0,2)$ supports that the enhancement in $I_{sw}(2,2)$ is originated from the CPS such that two electrons of the Cooper pair split into the two NWs.

\section*{Conclusion}
In summary, we examined CPS and LPT in a ballistic InAs DNW using a Josephson junction. We observed a large CPS efficiency for the DNW due to suppression of LPT into the respective NWs resulting from 1D e-e interaction. The CPS efficiency is tunable by adjusting the gate voltages and can well exceed 50 \% for narrow NW channels. Additionally the inter-wire superconducting gap is greater than the intra-wire one when each NW has a single channel. These results suggest that the InAs DNW coupled to a superconductor can hold time-reversal invariant topological superconductivity and Majorana Kramers pairs at the edges with no magnetic field. 
Our results pave the way for the utilization of 1D e-e interaction to reveal the physics of high-efficiency CPS and engineer MFs in double nanowire systems coupled to a superconductor via CPS.

\clearpage
\begin{figure}[t]
\centering
\includegraphics[width=0.55\linewidth]{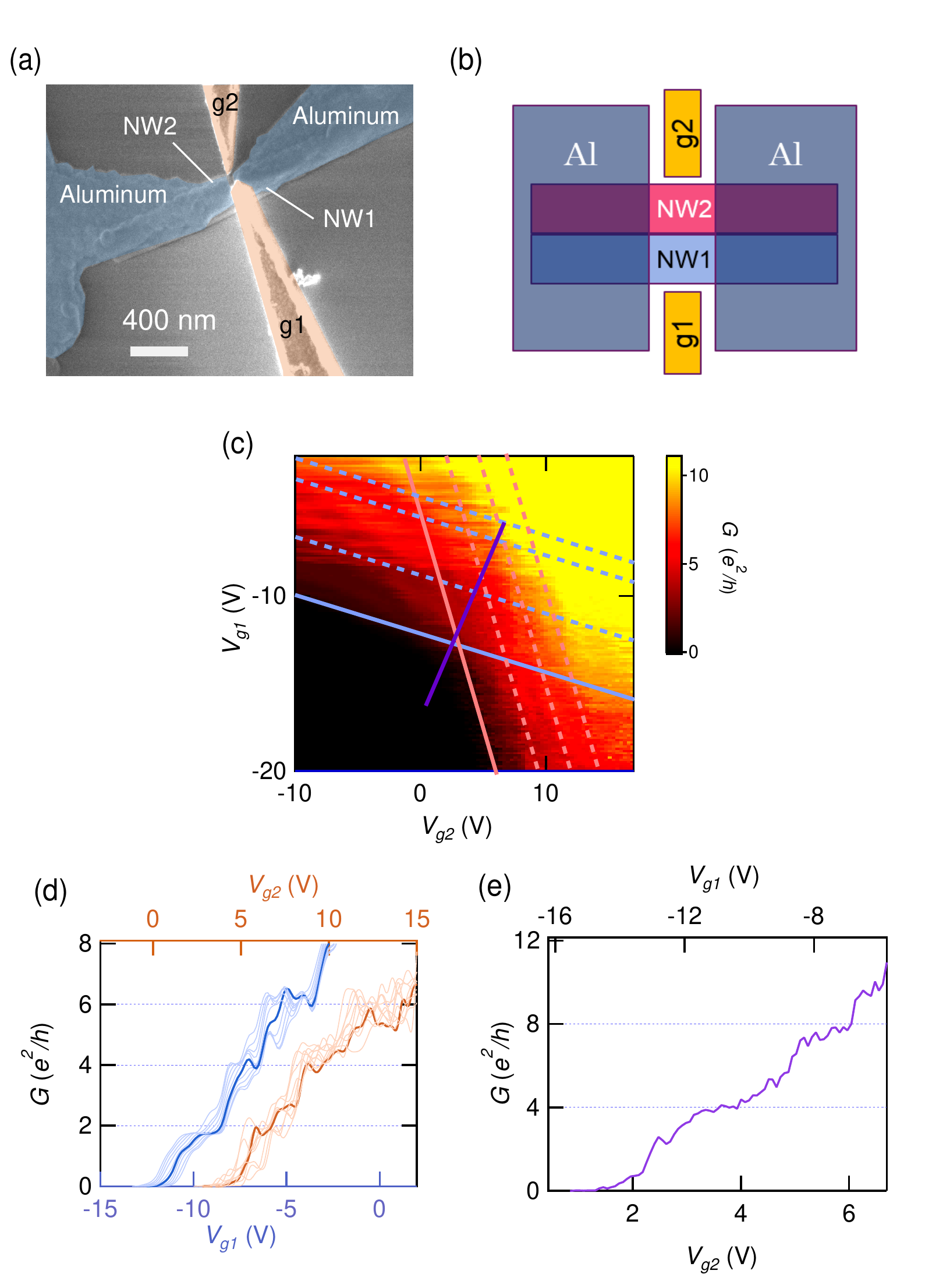}
\caption{{\bf Device structure and normal state conductance.}
(a) SEM image of the device. Two Al electrodes (blue) spaced by approximately 20 nm are placed on an InAs DNW. Two top gate electrodes (orange) spaced by approximately 80 nm are contacted to the DNW. The scale bar represents 400 nm.
(b) Schematic image of the device. NW1 and NW2 are mainly gated by electrode g1 with voltage $V_{g1}$ and electrode g2 with voltage $V_{g2}$, respectively.
(c) Differential conductance $G$ in units of $e^2/h$ as a function of $V_{g1}$  and $V_{g2}$ measured for magnetic field $B = 250$ mT and $50$ mK. The blue (red) solid line follows the NW1 (NW2) pinch-off points. The dashed lines parallel to the solid lines indicate transitions between the respective NW plateaus (see (d)). 
(d) The blue (red) lines indicate the NW1 (NW2) conductance measured by setting $V_{g2}$ between -5.0 and -8.0 V ($V_{g1}$ between -17.0 and -20.0 V). 
 $G$ against $V_{g1}$ (blue) measured by setting $V_{g2}$ between -5.0 and -8.0 V, where NW1 is pinched off, and $G$ against $V_{g2}$ (red) by setting $V_{g1}$ between -17.0 and -20.0 V, where NW2 is pinched off. All conductance curves show plateau-like features at 2, 4, and 6 $e^2/h$, as shown by the bold curves.
(e) $G$ plotted along the purple solid line in (c) where both NWs are equally populated. The conductance shows plateaus of 4 and 8 $e^2/h$ when both NWs have two and four propagating channels, respectively.
}
\label{fig1} 
\end{figure}

\begin{figure}[t]
\centering
\includegraphics[width=0.8\linewidth]{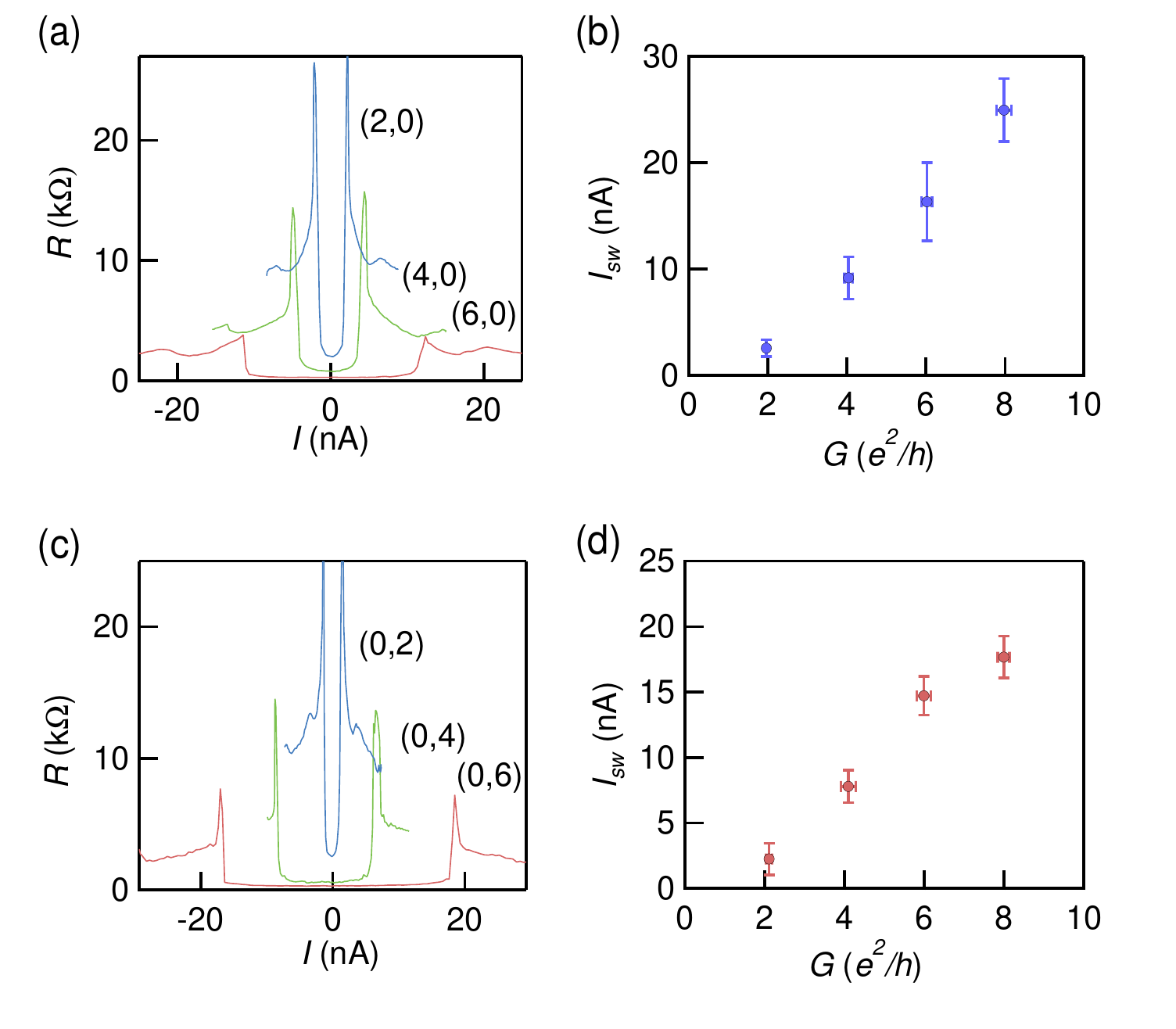}
\caption{{\bf Supercurrent due to local pair tunneling into each NW.}
(a) Typical differential resistance $R$ against bias current $I$ at $B = 0$ T measured in the conductance plateau regions of (2,0), (4,0), and (6,0), respectively, as shown in Fig. 1(d). The supercurrent flows in the Josephson junction in the region of $R\simeq 0\, {\rm \Omega }.~I_{sw}$ is evaluated from the peak position.
(b) $I_{sw}$ against $G$ derived from measurement results shown in (a). The bars indicate variations of $I_{sw}$ and $G$ in the measurement performed at various points of the respective plateaus. $I_{sw}$ monotonically depends on $G$.
(c) and (d) Identical plots to (a) and (b), respectively, but for the conductance plateau regions of (0,2), (0,4), and (0,6). 
}
\label{fig2} 
\end{figure}

\begin{figure}[t]
\centering
\includegraphics[width=0.7\linewidth]{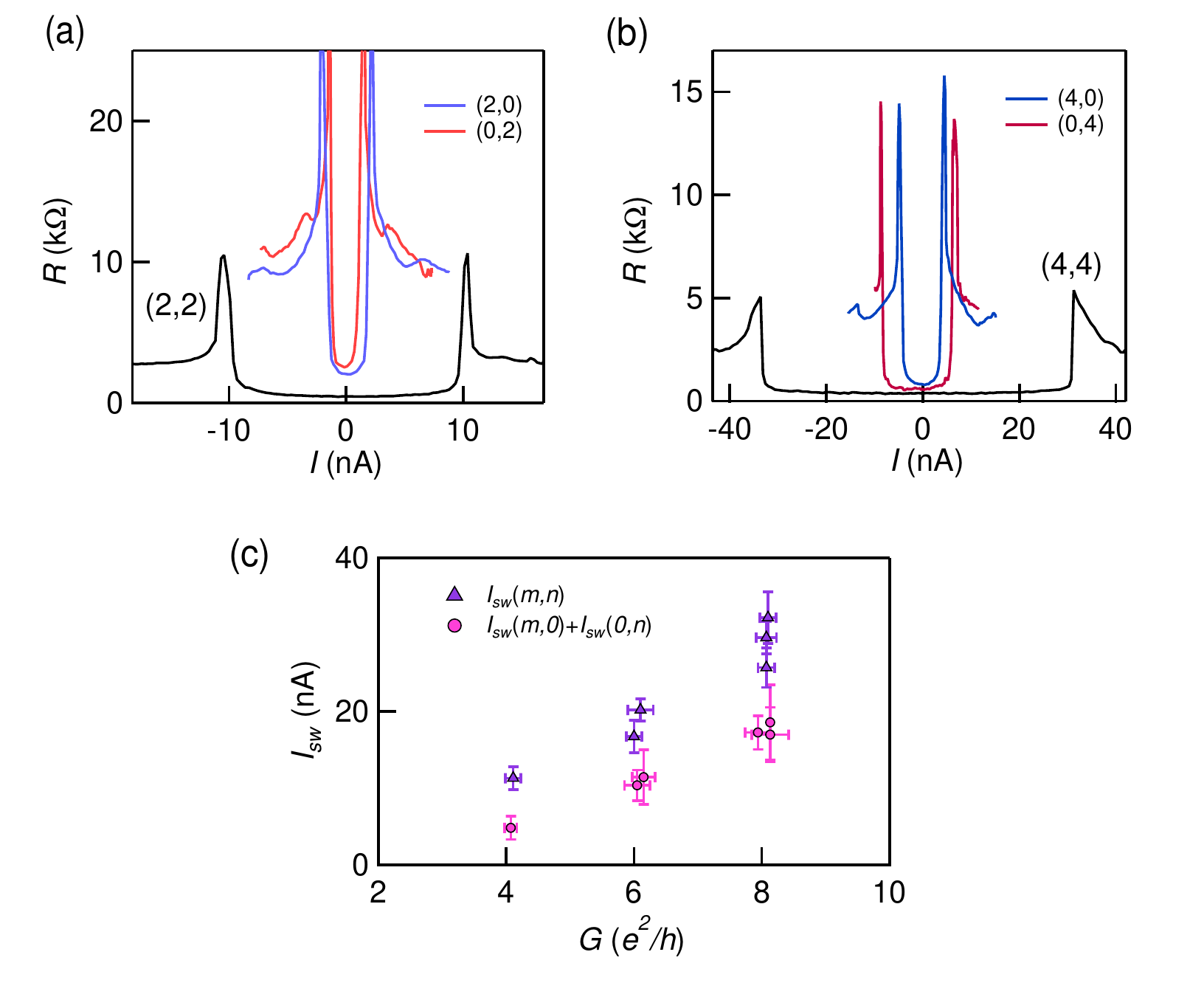}
\caption{{\bf Supercurrents in various conductance plateau regions.}
(a) Differential resistance $R$ against $I$ in the conductance plateau regions of (2,0), (0,2), and (2,2). $I_{sw}$ in the (2,2) region is much larger than the sum of the $I_{sw}$ values in the (2,0) and (0,2) regions.
(b) $R$ against $I$ in the conductance plateau regions of (4,0), (0,4), and (4,4). $I_{sw}$ in the (4,4) region is much larger than the sum of the $I_{sw}$ values in the (4,0) and (0,4) regions.
(c) $I_{sw}(m,n)$ against $G(m,n)$ in the conductance plateau regions $(m,n)=(2,2), (2,4), (4,2), (2,6), (6,2)$, and $(4,4)$, respectively, and the sum of $I_{sw}(m,0)$ and $I_{sw}(0,n)$ against the sum of $G(0,n)$ and $G(m,0)$ in the conductance plateau regions $(0,n)$ = (0,2) and (0,4) and $(m,0)$ = (2,0) and (4,0). The bars indicate variations of $I_{sw}$ and $G$ in the measurement performed at various points of the respective plateaus. $I_{sw}(m,n)$ is significantly larger than $I_{sw}(n,0)+ I_{sw}(0,m)$ because of the CPS contribution to the DNW.
}
\label{fig3} 
\end{figure}

\begin{figure}[t]
\centering
\includegraphics[width=0.75\linewidth]{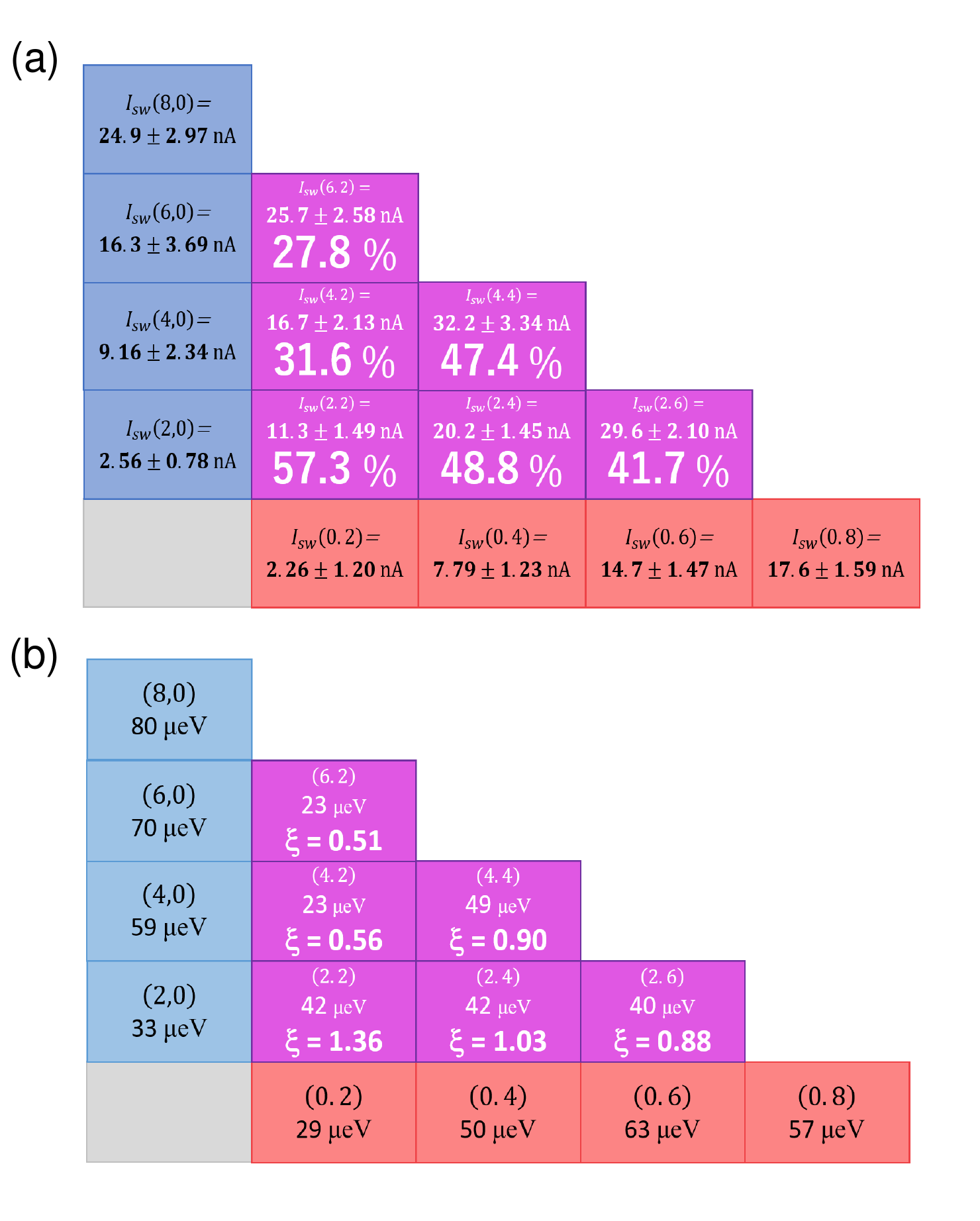}
\caption{{\bf CPS efficiency and gap energies of the inter-wire and intra-wire superconductivity}
(a) Schematic table of $I_{sw}$ and CPS efficiency $\eta$ obtained for various $m$ and $n$ values.
$I_{sw}$ enhancement due to CPS is observed for all conductance plateaus in the DNW regions. The CPS $\eta$ is significantly larger than 50 \% in the (2,2) region.
(b) The estimated superconducting gap energies and the ratio of the inter-wire and intra-wire superconductivity $\xi$ in the respective ($m,n$) regions. $\xi$ is larger than unity in the (2,2) and (2,4) regions.
}
\label{fig4} 
\end{figure}

\begin{figure}[t]
\centering
\includegraphics[width=0.75\linewidth]{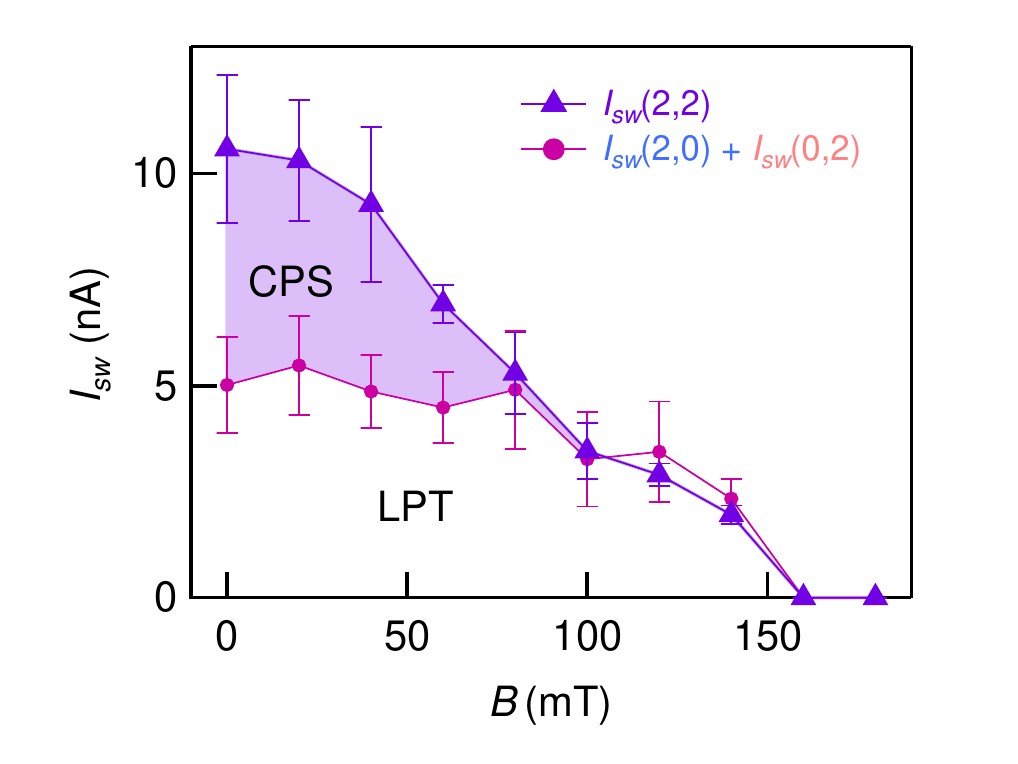}
\caption{{\bf Magnetic field dependence of CPS and LPT components.}
$I_{sw}(2,2)$ and $I_{sw}(2,0)+I_{sw}(0,2)$ measured at various magnetic fields of $B = 0$ to $180$ mT. $I_{sw}(2,2)$ arises from both LPT into separate NWs and CPS into both NWs. The purple-shaded region corresponds to the $I_{sw}$ enhancement due to CPS. The CPS component gradually decreases and vanishes at $B=80$ mT, whereas the LPT component is unchanged up to $B=80$ mT and then decreases down to $I_{sw}$ = 0 nA at $B=160$ mT.
 }
\label{fig5} 
\end{figure}

\clearpage

\bibliographystyle{Science}

\section*{Acknowledgments}
We thank P. Stano and C.-H. Hsu for their fruitful discussion. 
Funding: This work was partially supported by a Grant-in-Aid for Scientific Research (B) (No. JP18H01813), 
a Grant-in-Aid for Young Scientific Research (A) (Grant No. JP15H05407), 
a Grant-in-Aid for Scientific Research (A) (Grant No. JP16H02204), 
a Grant-in-Aid for Scientific Research (S) (Grant No. JP26220710), 
JSPS Early-Career Scientists (No. JP18K13486), the JSPS Program for Leading Graduate Schools (MERIT) from JSPS, 
Grants-in-Aid for Scientific Research on Innovative Area Nano Spin Conversion Science (Grants No. JP17H05177), 
a Grant-in-Aid for Scientific Research on Innovative Area Topological Materials Science (Grant No. JP16H00984) from MEXT, 
JST CREST (Grant No. JPMJCR15N2), 
the ImPACT Program of Council for Science, Technology, and Innovation (Cabinet Office, Government of Japan), 
the Ministry of Science and Technology of China (MOST) through the National Key Research and Development Program of China (Grant Nos. 2016YFA0300601, 2017YFA0303304), 
the National Natural Science Foundation of China (Grant Nos. 91221202, 91421303), 
and the Swedish Research Council and NanoLund (VR). 

\section*{Author contributions} S.M. and S.T. conceived the experiments. K.L., S.J., L.S., and H.X. grew the NWs. K.U. fabricated the device and S.M., S.B., H.K., Y.S., and Y.T. contributed to the fabrication. K.U. and S.M. executed the measurements. K.U., S.M., and S.T. analyzed and interpreted the data and wrote the paper. S.T. supervised the study.

\clearpage

\section*{Supplementary materials}
Supplementary Note 1 to 7\\
Figs. S1 to S8\\
References \textit{(1-5)}
\clearpage
\end{document}